\documentclass[pra,aps,superscriptaddress,twocolumn,floatfix]{revtex4-1}

\usepackage{dsfont}
\usepackage{mathtext}
\usepackage[cp1251]{inputenc}
\usepackage[english, russian]{babel}

\usepackage{mathtools}
\usepackage[usenames,dvipsnames]{xcolor}
\usepackage{amsmath}
\usepackage{amssymb,mathrsfs}
\usepackage{graphicx}

\usepackage{mathtext}

\makeatletter
\renewcommand\@make@capt@title[2]{%
    \@ifx@empty\float@link{\@firstofone}{\expandafter\href\expandafter{\float@link}}%
        {\textbf{#1}}\@caption@fignum@sep#2\quad}%
\makeatother

\makeatletter
\renewcommand{\fnum@figure}{\textbf{Fig.~\thefigure}}
\makeatother

\newcommand{\om}{\omega}

\newcommand{\prt}{\partial}

\makeatletter

\renewcommand*{\p@subsection}{}

\renewcommand*{\p@subsubsection}{}
\makeatother

\graphicspath{{Figures/}}

\begin{document}

\title{Evolution of intensive light pulses in a nonlinear medium with the Raman effect}



\author{
S. K. Ivanov, A. M. Kamchatnov\\
~\\
\small{\it Institute of Spectroscopy, Russian Academy of Sciences, Troitsk, Moscow, 108840, Russia}\\
\small{\it Moscow Institute of Physics and Technology, Institutsky lane 9, Dolgoprudny, Moscow region, 141700, Russia}}

\begin{abstract}
In this paper, we study the evolution of intensive light pulses
in nonlinear single-mode fibers. The dynamics of light in such
fibers is described by the nonlinear Schr\"odinger equation
with the Raman term, due to stimulated Raman self-scattering.
It is shown that dispersive shock waves are formed during the
evolution of sufficiently intensive pulses. In this case the situation
is much richer than for the nonlinear Schr\"odinger
equation with Kerr nonlinearity only. The Whitham equations are
obtained under the assumption that the Raman term can be considered
as a small perturbation. These equations describe slow evolution
of dispersive shock waves. It is shown that if one takes
into account the Raman effect, then dispersive shock waves
can asymptotically acquire a stationary profile.
The analytical theory is confirmed by numerical calculations.

~

\textit{Keywords:} nonlinear medium, Raman effect, dispersive shock waves, solitons, Whitham modulation equations.
\end{abstract}



\maketitle

\section{Introduction}\label{sec1}

The problem of evolution of light pulses in wave-guides is a
subject of active modern experimental and theoretical research.
It is well known that if one neglects the effects of dissipation and dispersion,
then the theory of nonlinear propagation
of light envelope suffers from a wave breaking singularity developed at some
finite fiber length after which
a formal solution of nonlinear wave equations becomes
multi-valued and loses its physical meaning.
The account of dispersion eliminates such a non-physical behavior,
but the evolution equations for the envelopes acquire
higher order derivatives which makes their analytical treatment more difficult.
Nevertheless, the following qualitative picture
of the phenomenon can be inferred from numerical experiments.
After the wave breaking moment, instead of multi-valued region,
an expanding region of fast nonlinear oscillations is formed.
Envelope parameters in such a structure change slowly compared with
the characteristic oscillations frequency and their wavelength.
This region of fast oscillations is called ``dispersive shock wave'' (DSW).
In nonlinear optics, such structures were observed long ago
(see, for instance, \cite{tsj-85,rg-89}).
But even earlier this phenomenon had been studied in the water waves dynamics
(see, for instance, \cite{BenLigh-1954})
and plasma physics (see, for instance, \cite{tbi-70}).
The general nature of this phenomenon, which arises as a
result of interplay of nonlinear and dispersion
effects with low viscosity, was studied by R.~Z.~Sagdeev
\cite{Sagdeev-1964}.
The introduction of small dissipation can, in some cases,
balance the dispersive effects so that the DSW
eventually acquires a steady profile, but remains oscillatory in
space.
The simplest and, apparently, most powerful theoretical
approach to a description of DSWs was formulated by
Gurevich and Pitaevskii \cite{GP-1973} in framework of the Whitham
theory of modulation of nonlinear waves \cite{whitham-65,whitham-77} which was
based on large difference between scales of the wavelength
of nonlinear oscillations within DSW and the size of the
whole DSW. To date, the theory and experimental
investigations of DSWs have been greatly developed,
It has spread to other fields of nonlinear physics,
including the dynamics of nonlinear waves in a
Bose-Einstein condensates; see, for instance, the review article
\cite{ElHoefer-2016} and references therein.
In particular, the theory of DSWs has been applied
to the nonlinear Schr\"odinger (NLS) equation
for media with normal dispersion and Kerr nonlinearity
\cite{gk-87,eggk-95}. And it agrees perfectly well
with experimental studies of the evolution of specially formed
rectangular pulses in optical fibers \cite{Xu-2017}.

The NLS equation theory is one of the main ways of
describing waves in nonlinear optics and it
usually gives a good qualitative explanation of
the main features of the phenomenon.
However, it often turns out to be insufficient
for quantitative interpretation of the emerging DSWs.
Sometimes even small corrections to the NLS equation,
caused by taking into account small additional effects,
lead to a sufficiently radical deviations for long evolution
from predictions of the NLS equation theory.
As was demonstrated in \cite{Zakh-1971}, the NLS equation belongs
to a special class of so-called completely integrable equations.
Going beyond this class makes the inverse scattering method
used in the theory of the NLS equation inapplicable.
For instance, optical shock waves were observed in intensive
light beams propagating in photorefractive crystals
with defocusing saturated nonlinearity \cite{wjf-2007},
so by increasing the intensity of the wave, the relative
role of nonlinearity decreases, which is not considered in the NLS theory.
In this case, the Whitham modulation equations are too
complicated for finding their global analytical solution.
In spite of that, for the special case of the initial condition
in the form of an intensity step-like discontinuity, the method of the work
\cite{El05} gives the main characteristics of DSW \cite{egkkk-07}.
A similar theory can be developed for optical
DSW in colloidal media \cite{ans-17}.
A more general form of initial pulses can be considered with the use of
the recently developed method \cite{Kamch-2019}.

Changing the type of nonlinearity can lead not only to quantitative
differences from the NLS theory but also to qualitatively new effects.
For instance, the delay of the nonlinear response of the medium to the wave
field turns the NLS equation into so-called
``derivative nonlinear Schr\"odinger equation'',
which contain in addition to the Kerr-type nonlinearity its time derivative.
Such a modification of the NLS equation radically changes the theory of DSWs.
As a result, new wave structures of the combined type become possible.
For example, the oscillation region can be combined with rarefaction waves,
as well as other wave configurations (see \cite{IvKamch-2017}) can be
observed.

Other qualitatively new effects are caused by small perturbations
of dissipative type.
These perturbations at a sufficiently large time become
comparable with small modulation of the wave packet, which can
lead to the stabilization of DSW so that it acquires a stationary profile.
This phenomenon can also occur when one takes into account the Raman
effect in the propagation of light pulses in the wave-guides.
As was mentioned in \cite{Kivshar-1990}, in the small-amplitude approximation,
the NLS equation with an additional Raman term can be reduced to the
Korteweg-de Vries-Burgers (KdV-B) equation.
The presence of stationary DSWs for the KdV-B equation was
established in \cite{johnson-70} by the direct perturbation
theory (\cite{gp-87,akn-87}) and by using the
Whitham method (see also \cite{Kamchatnov-2016}).
However, taking into account perturbing terms of this type
in the NLS theory is much more complicated (\cite{Kamch-2004,lpk-12})
and therefore requires separate consideration.
The aim of this work is to study the impact of
the Raman effect, which is caused by the influence
of retardation of the fiber material response to variations of the electromagnetic signal
on the dynamics of light pulses propagating in a single-mode fiber.
First, in section \ref{sec3}, we consider the influence
of this effect on the dynamics of linear waves
that propagate along a uniform background.
In section \ref{sec4} we find the leading
dispersion and nonlinear corrections
to the dispersionless linear propagation
of disturbances since some qualitative
features of the behavior of a light pulse envelope
in fiber can be already explained
by the small-amplitude limit of the evolution equations.
Then we proceed to the description of several stages
in the DSW evolution after
wave breaking.
As will be shown, in the first stage,
when the length of light propagation
through the fiber is sufficiently small,
the Raman effect can be neglected.
Then the system can be described by the ordinary NLS equation
with Kerr nonlinearity.
The solution of the Whitham equations for
this equation in some characteristic cases is well known.
At the next stage, the Raman effect becomes efficient.
In this case, the evolution of the DSW is described
by the perturbed Whitham equations, which
will be derived in section \ref{sec5}
within the framework of the theory developed
in \cite{Kamch-2004}.
We will show that this effect affects differently on
the waves that propagate in different directions.
Finally, it will be shown that the shock wave
directed in the positive direction of the time
axis $t$ becomes stationary, while the parameters
of the DSW propagating in the opposite direction
continue to evolve with increasing amplitude of the
DSW and the duration of the wave structure.
The analytical results obtained in the work are
confirmed by numerical calculations.

\section{The model}\label{sec2}

We proceed from a standard approach
(see, for instance, \cite{KivsharAgrawal-2003}),
in which the dynamics of the electric field envelope
$E(x,t)$ of the light wave is described by the NLS equation
taking into account normal dispersion and defocusing
Kerr nonlinearity, whereas we neglect attenuation:
\begin{equation} \label{NLS_Dimention}
    \frac{\partial E}{\partial X} + \beta_1 \frac{\partial E}{\partial T} +
    \frac{i}{2}\beta_2\frac{\partial^2 E}{\partial T^2} - i\tilde{\gamma}|E|^2E= 0,
\end{equation}
where $X$ is a coordinate along the waveguide, $T$ is a time,
$\beta_1$ is reverse group velocity of the waves ($v_{gr}=1/\beta_1$)
and $\beta_2$ is the parameter that determines pulse expansion,
$\tilde{\gamma}$ is non-linear coefficient which determined by the expression
\begin{equation*} \label{}
    \tilde{\gamma} = \frac{n_2 \omega_0}{c_{l}E_{eff}}, \qquad E_{eff} = \pi w^2
\end{equation*}
for the carrier frequency $\omega_0$.
Here $w$ is the parameter of the Gaussian mode,
$n_2$ is the nonlinear refractive index, $c_{l}$ is the light velocity in vacuum.
Thus, the second term in the evolution equation
(\ref{NLS_Dimention}) describes wave transfer
with group velocity, and the last term corresponds to Kerr nonlinearity.
The term with the positive coefficient $\beta_2>0$ is a
quadratic dispersion. Equation (\ref{NLS_Dimention}) by substitution
\begin{equation*} \label{}
    x=-\tilde{\gamma}I_0X, \quad t=\sqrt{\frac{\tilde{\gamma}I_0}{\beta_2}}(T-\beta_1 X),
    \quad q = \frac{E}{\sqrt{I_0}},
\end{equation*}
where $ I_0 $ is the characteristic intensity of the system,
is converted to a conventional dimensionless form
\begin{equation} \label{NLS_Dimentionless}
    iq_x + \frac12 q_{tt} - |q|^2q = 0.
\end{equation}

As mentioned in the Introduction, the propagation
of pulses along sufficiently long fibers is significantly
affected by small effects that are not taken into account
in the NLS approximation, such as higher-order dispersion,
self-steepening and Raman scattering (Raman effect).
The influence of the self-steepening effect on the DSW evolution
was discussed in detail in \cite{IvKamch-2017},
where it was shown that it leads to the formation of complex combined structures.
The Raman effect describes the mixing of the frequency
of stimulated Raman self-scattering. And its consideration leads
to an additional term in the evolution equation so that the
equation for the light pulse envelope in dimensionless
variables has a form
\begin{equation} \label{NLS_Raman}
    iq_x + \frac12 q_{tt} - |q|^2q + \gamma q(|q|^2)_t= 0.
\end{equation}
Here $\gamma$ is a constant, which characterizes the slope
of the SRS-gain line. It is usually a small parameter of
the system, which allows us to consider the last term
of the equation as a perturbation for the description
of DSW in the Whitham theory.
It is worth noting that the manifestations of
the self-steepening and the Raman effect are quite different and therefore they
can be identified separately.

We shall start with the study of linear waves
which propagate along a uniform wave background.

\section{Linear waves}\label{sec3}

Let the light pulse propagate along a uniform wave
background with the amplitude $\sqrt{I_0} = |q_0| = \mathrm{const}$.
To see the role of the Raman term, we find the solution
of the linearized NLS equation with the Raman term for evolution of
a pulse in a linear approximation.
It is convenient to make a substitution
$q(x,t)=\tilde{q}(x,t)\exp{(-{i}I_0 x)}$.
Then equation (\ref{NLS_Raman}) takes the form
\begin{equation} \label{NLS_transformed}
    i\tilde{q}_x+\frac12\tilde{q}_{tt}+(I_0-|\tilde{q}|^2)\tilde{q}+\gamma \tilde{q}(|\tilde{q}|^2)_t=0.
\end{equation}
This replacement does not change the properties
of the equation since the phase of the wave is
determined up to a constant. The evolution of a small
perturbation $\delta q$ propagating along a homogeneous background,
\begin{equation} \label{Perturbation}
    \tilde{q}=\sqrt{I_0}+\delta q,\qquad |\delta q|\ll\sqrt{I_0},
\end{equation}
can be described by a linearized equation
\begin{equation} \label{NLS_linearized}
    i\delta q+\frac12\delta q_{tt}-I_0(\delta q+\delta q^*)+\gamma I_0(\delta q+\delta q^*)_t=0
\end{equation}
with the initial condition $\delta {q}|_{x=0}=\delta {q_0}(t)$.
After separation of real and imaginary parts
\begin{equation} \label{Re_Im}
    \delta q=A+iB
\end{equation}
we obtain from (\ref{NLS_linearized})
\begin{equation} \label{Eq_For_A_B}
    A_x+\frac12 B_{tt}=0, \quad B_x-\frac12 A_{tt}+2I_0A-2\gamma I_0A_t=0.
\end{equation}
The function $B$ can be excluded, and then from this system
we get a linear equation for $A$:
\begin{equation} \label{Eq_A}
    A_{xx}-I_0A_{tt}+\gamma I_0A_{ttt}+\frac14 A_{tttt}=0.
\end{equation}
This equation can be solved by the Fourier method. To this
end, we note that linear harmonic waves $A\propto\exp{[{ i}(k x-\omega t)]}$
satisfy the dispersion law
\begin{equation} \label{Dipersion_Law}
    k=\pm k(\omega), \quad k(\omega)=\omega\sqrt{\frac{\omega^4}{4}+I_0(1+i\gamma\omega)}.
\end{equation}
The imaginary unit in the dispersion law means that for $ \gamma \neq0 $
there is a damping or amplification of linear waves. General solution of
the equation (\ref{Eq_A}) can be written as
\begin{equation} \label{General_solution_for_A}
\begin{split}
    A(x,t)= & \int_{-\infty}^{+\infty}W_1(\omega)e^{i(k(\omega) x-\omega t)}\frac{d\omega}{2\pi} \\
            & +\int_{-\infty}^{+\infty}W_2(\omega)e^{i(-k(\omega) x-\omega t)}\frac{d\omega}{2\pi},
\end{split}
\end{equation}
where functions $W_{1,2}(\omega) $ are determined from the initial conditions
\begin{equation} \label{Initial_condition_for_W}
\begin{split}
    A(0,t) & = \int_{-\infty}^{+\infty}[W_1(\omega)+W_2(\omega)]
    e^{-i\omega t}\frac{d\omega}{2\pi}  \\
    A_x(0,t) & = i\int_{-\infty}^{+\infty}k(\omega)[W_1(\omega)
    -W_2(\omega)]e^{-i\omega t}\frac{d\omega}{2\pi}.
\end{split}
\end{equation}
After standard calculations we arrive at the solution
expressed in terms of the Fourier transform $\widehat{I'_0}(\omega)$ and
$\widehat{\varphi'_0}(\omega)$ of the initial (input)
intensity $I'(x,t)=2\sqrt{I_0}A(x,t)$ and phase $\varphi'(x,t)=B(x,t)/\sqrt{I_0}$ disturbances,
\begin{equation} \label{}
\begin{split}
    W_1(\omega) & = \frac{1}{4\sqrt{I_0}}\left[ \widehat{I'_0}(\omega)
    -i\frac{\omega^2}{k(\omega)}I_0\widehat{\varphi'_0}(\omega) \right], \\
    W_1(\omega) & = \frac{1}{4\sqrt{I_0}}\left[ \widehat{I'_0}(\omega)
    +i\frac{\omega^2}{k(\omega)}I_0\widehat{\varphi'_0}(\omega) \right].
\end{split}
\end{equation}
Suppose that at the initial moment the phase of
the wave is constant and there is only the
intensity perturbation, that is, $ \varphi'_0 = 0$, then
\begin{equation} \label{Intensity_perturbation}
\begin{split}
    I'(t,x)=\frac{1}{4\pi}\int_{-\infty}^{+\infty} \widehat{I'_0}(\omega)
    \left[e^{ixf_1(\omega)}+e^{ixf_2(\omega)} \right] d\omega,
\end{split}
\end{equation}
where
\begin{equation} \label{}
    f_1(\omega)=k(\omega)-\omega \frac{t}{x}, \quad
    f_2(\omega)=-k(\omega)-\omega \frac{t}{x}.
\end{equation}
These integrals can be estimated for a large distance of
propagation $x$ by the method of stationary phase resulting in
\begin{equation} \label{Approximate_Lin_Left}
    \delta I(t,x) \simeq \frac{2\widehat{I'_0}(\om^{(1)}_0)}{\sqrt{2 x
    {|\frac{d^2f_1}{d\om^2}|}_{\om_0^{(1)}}}}
    \cos{ \left (xf_1(\om^{(1)}_0)-\frac{\pi}{4} \right )}
\end{equation}
for the wave that propagates in the negative direction of the $t$-axis, and
\begin{equation} \label{Approximate_Lin_Right}
    \delta I(t,x) \simeq \frac{2\widehat{I'_0}(\om^{(2)}_0)}{\sqrt{2 x
    {|\frac{d^2f_2}{d\om^2}|}_{\om_0^{(2)}}}}
    \cos{ \left (xf_2(\om^{(2)}_0)-\frac{\pi}{4} \right )}
\end{equation}
for a linear wave propagating to the right.
Here $\omega^{(1)} _0$ and $\omega^{(2)}_0$ are the values of
$\omega$ at the points of
the stationary phase that are defined by the equations
\begin{equation} \label{}
    \frac{df_1}{d\omega}=0, \qquad \frac{df_2}{d\omega}=0.
\end{equation}

\begin{figure}[t] \centering
\includegraphics[width=8cm]{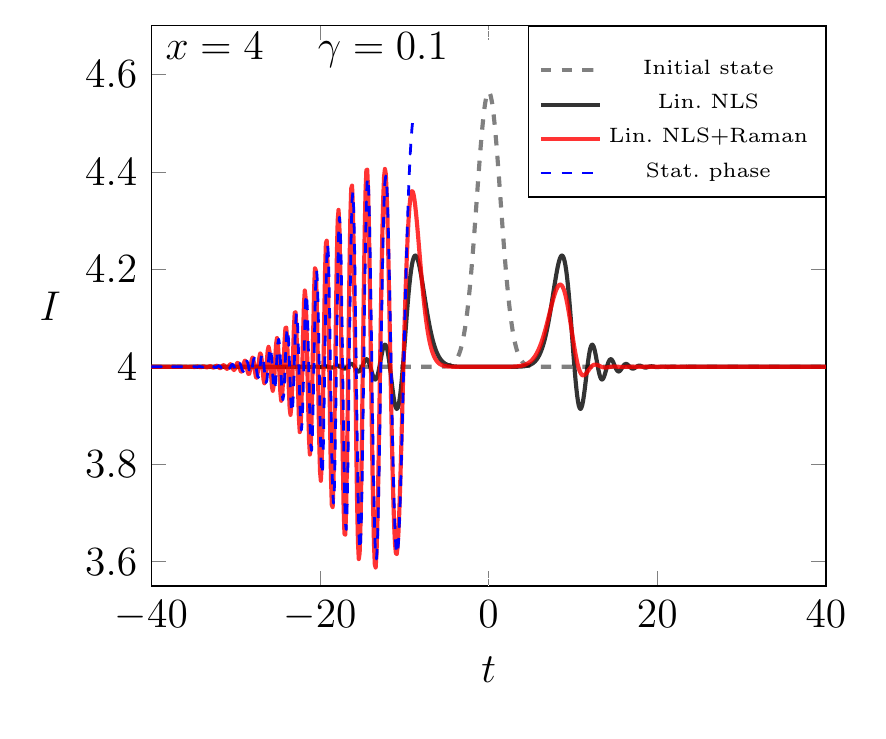}
\caption{Evolution of the pulse in the linear approximation of
the NLS equations with Raman term (\ref{NLS_Raman}) with $\gamma=0.1$.
The dashed gray line shows the initial perturbation, the dashed blue
line corresponds  to the
stationary phase approximation (\ref{Approximate_Lin_Left}) and
(\ref{Approximate_Lin_Right}), and the red solid line corresponds to
the numerical calculation of the integral (\ref{Intensity_perturbation})
for $x=4$ with the initial perturbation (\ref{Initial_Dist}) with $a=1$.
Black solid line shows linear waves obeying a linearized equation
(\ref{NLS_linearized}) with $ \gamma = 0 $.}
\label{Fig1}
\end{figure}
In figure~\ref{Fig1} we compare the numerical calculations of the
integral (\ref{Intensity_perturbation}) with its approximate
estimates (\ref{Approximate_Lin_Left}) and (\ref{Approximate_Lin_Right})
for the initial perturbation
\begin{equation} \label{Initial_Dist}
     I'_0(t)=\frac{1}{\sqrt{\pi}a}\exp{\left(-\frac{t^2}{a^2}\right)},\quad
     \widehat{I'_0}(\omega)=\exp{\left(-\frac{\omega^2a^2}{4}\right)}.
\end{equation}
As we see, the pulse splits into two smaller pulses, however,
on the contrary to the NLS case, they are
not symmetrical pulses propagating in opposite directions.
Now, these two pulses have different profiles.
It can be seen that the pulse propagating in
the positive direction of the $t$-axis attenuates,
while the impulse propagates in the negative direction is amplified.
This is the manifestation of lack of the time inversion invariance,
which is caused by the last term in (\ref{NLS_Raman}).
The approximate solution (\ref{Approximate_Lin_Left}),
(\ref{Approximate_Lin_Right}) shows that the amplitude
of a linear wave packet propagating to the left
increases monotonically. However, one should not forget
that this theory is valid for small deviations from the background intensity.
It should be noted that the asymptotic solution (\ref{Approximate_Lin_Left}) and (\ref{Approximate_Lin_Right})
describes well the wave packet even for not very large $x$.

\begin{figure}[t] \centering
\includegraphics[width=8cm]{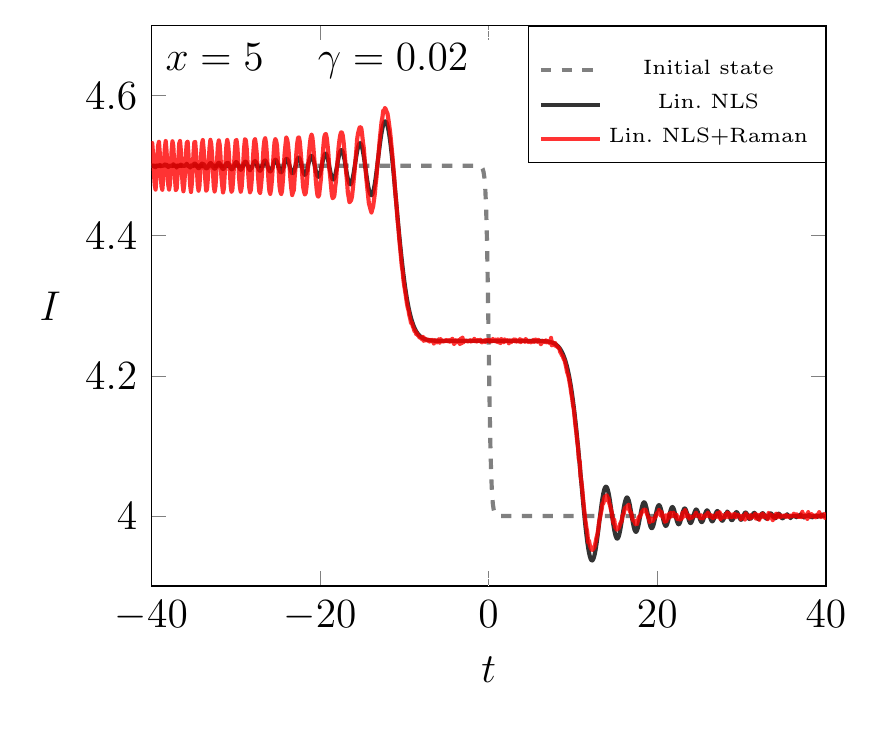}
\caption{Evolution of the pulse in the linear approximation of the NLS equation
with Raman term (\ref{NLS_Raman}) with $\gamma=0.02$.
The dashed gray line shows the initial perturbation as an initial discontinuity
and the red solid line corresponds to the numerical calculation of the integral
(\ref{Intensity_perturbation}) for $x=5$ with the initial perturbation
(\ref{Initial_Dist_tanh}) with $I^L=0.5$, $I^R=0$ and $\Delta=0.3$.
The black solid line shows the linear waves obeying the linearized equation
(\ref{NLS_linearized}) with $\gamma=0$.}
\label{Fig2}
\end{figure}

The figure~\ref{Fig2} shows the dependence
of linear wave intensity on time
for the step-like initial condition,
which we model by the formula
 \begin{equation} \label{Initial_Dist_tanh}
     I'_0(t)=\frac{1}{2}\left\{I^R+I^L+(I^R-I^L)\tanh{\left(\frac{t}{\Delta}\right)}\right\},
\end{equation}
where ``step width'' $\Delta$ should be taken small
enough so that
its effect disappears at sufficiently large time.
The constants $I^R$ and $I^L$ are responsible for the boundary values
of the light intensity on the right and left sides of discontinuity, respectively.
The problem of the evolution of the initial discontinuity
is one of the fundamental importance in the DSW theory,
and we will consider it in detail below.
Here we note that the difference between linear waves
at $\gamma=0$ and at $\gamma\neq0$ is
noticeable even at small values of $\gamma$,
especially for the left wave.

Let us turn now to the study of the effect of small
dispersion and nonlinearity on the light pulse evolution.

\section{Small-amplitude and weak dispersion limits}\label{sec4}

Let us turn to the study of the Raman effect influence
on the evolution of a DSW in the small-amplitude and weak dispersion limit
when they are taken into account in the main approximation.
That is, we are interested in the leading dispersion
and nonlinear corrections to the dispersionless
linear propagation of disturbances along the uniform background $ I_0 $.
To this end,
it is convenient to use the physical variables of intensity
$ I (x, t) $ and chirp $ u (x, t)$.
To go to the equations for these variables, we apply the Madelung transform
\begin{equation}
    q(x,t) = \sqrt{I(x,t)}\exp{\left({ i}\int^tu(x,t')dt'\right)}.
\end{equation}
After its substitution into Eq.~(\ref{NLS_Raman}),
separation of the real and imaginary parts
and differentiation of one of the equations with respect to $x$,
we get the system
\begin{equation} \label{Hydro}
\begin{split}
    & I_x+(uI)_t=0, \\
    & u_x+uu_t+I_t+\left(\frac{I^2_t}{8I^2}-\frac{I_{tt}}{4I}\right)_t=\gamma I_{tt}.
\end{split}
\end{equation}
The last term on the left-hand side of the second equation
describes the dispersion, and the term on the right-hand side of this
equation resembles the well-known Burgers viscosity
in the theory of water waves and other continuous media \cite{whitham-77}.

Using the last system and applying the standard perturbation
theory for the amplitude of the perturbation
and for the weak dispersion (see, for example, \cite{kamch-2000}),
one can obtain a small-amplitude
analog of the equation (\ref{NLS_Raman}). Let the wave propagate
in the positive
direction of the $t$-axis. Then an approximate equation for
$ I '= I-I_0 $ takes the form
\begin{equation} \label{eq1-ak}
\begin{split}
    \frac{\prt I'}{\prt x} +\sqrt{I_0}\frac{\prt I'}{\prt t} & +
    \frac{3}{2\sqrt{I_0}}I'\frac{\prt I'}{\prt t} \\
    & - \frac{1}{8\sqrt{I_0}}\frac{\prt^3 I'}{\prt t^3}
    = \frac{1}{2}\gamma\sqrt{I_0}\frac{\prt^2 I'}{\prt t^2}.
  \end{split}
\end{equation}
This is the Korteweg-de Vries--Burgers equation (KdV--B).
As is known (see \cite{gp-87, Kamchatnov-2016}),
the evolution of waves obeying this equation
is divided into three stages. In the first stage, when $x\ll\gamma^{-1}$,
the term that characterizes low viscosity has little effect
on the evolution of DSW and therefore it is neglected.
Then, in the second stage,
when $ x \sim \gamma^{-1} $, the influence of viscosity
becomes comparable to the modulation of DSW.
Obviously, in this case, viscosity cannot be neglected,
and then perturbation theory is used to
describe the DSW because of the smallness of the
viscosity coefficient. Finally, at the third stage,
at $ x \gg \gamma^{-1} $, the DSW becomes
stationary in $ x $, that is the output signal profile
ceases to depend on the fiber length. Such a behavior of the
DSW should be expected in the case of the equation (\ref{NLS_Raman})
for the wave propagating to the left.

However, the situation changes drastically for waves propagating
in the opposite direction. For a perturbation of intensity
$ I'$, the limiting equation is written in the form
\begin{equation} \label{eq2-ak}
\begin{split}
    \frac{\prt I'}{\prt x} -\sqrt{I_0}\frac{\prt I'}{\prt t} & +
    \frac{1}{\sqrt{I_0}}I'\frac{\prt I'}{\prt t} \\
    &  \frac{1}{8\sqrt{I_0}}\frac{\prt^3 I'}{\prt t^3}
    = -\frac{1}{2}\gamma\sqrt{I_0}\frac{\prt^2 I'}{\prt t^2}.
  \end{split}
\end{equation}
This is also the KdV--B equation, but now it has opposite signs
before the dispersion and viscosity.
This means that the orientation of the DSW is changed,
and wave amplification is expected instead of the standard
attenuation of the wave.

This difference between the waves oppositely directed
is caused by the absence of symmetry of the equation (\ref{NLS_Raman})
with respect to the replacement $ t \rightarrow-t $. Obviously,
this property will affect the evolution of the pulse described by
the complete equation (\ref{NLS_Raman}). Let us turn to the
description of the formation of DSWs and the derivation of
equations describing their dynamics.

\section{Dispersive shock wave formation}\label{sec5}

We first discuss the dispersionless limit of hydrodynamic
equations that follows from the NLS equation
(\ref{NLS_Dimentionless}) when the dispersion is neglected.
In this limit, equations (\ref{Hydro}) without the Raman term
is transferred to the system
\begin{equation} \label{Hydro_Dispersionless}
\begin{split}
    & I_x+(uI)_t=0, \\
    & u_x+uu_t+I_t=0,
\end{split}
\end{equation}
where the first equation can be interpreted as the
continuity equation for the intensity
$I$ and the second one as the Euler equation for the
``flow velocity'' $u$. This system can
be cast in a standard way to the Riemann diagonal form
\begin{equation} \label{}
    \frac{\prt r_{\pm}}{\prt x}+\frac1{v_{\pm}}\frac{\prt r_{\pm}}{\prt t}=0
\end{equation}
for the Riemann invariants
\begin{equation} \label{}
    {r_{\pm}}=\frac{u}2\pm\sqrt{I}
\end{equation}
with inverse velocities
\begin{equation} \label{}
    \frac1{v_{\pm}}=u\pm\sqrt{I}.
\end{equation}
The functions introduced here will be needed to select the initial
conditions corresponding to the wave propagating in a certain direction.

Let the initial (input) conditions have a step-like form
\begin{equation}\label{Initial_Step}
\begin{split}
    I(x=0)=
    \begin{cases}
        I^L, & \quad t<0, \\
        I^R, & \quad t>0,
    \end{cases}\quad \\
        u(x=0)=
    \begin{cases}
        u^L, & \quad t<0, \\
        u^R, & \quad t>0.
    \end{cases}\quad
\end{split}
\end{equation}
Under these initial conditions, the wave will break at $x=0$.
If the evolution of a light envelope $q (x,t)$ is
described by the equation (\ref{NLS_Dimentionless}),
then during the propagation two characteristic structures would be formed:
rarefaction wave and DSW. From the form of hydrodynamic
equations (\ref{Hydro}) and the small-amplitude limit, it is clear that
in the general case the same structures will be formed. However, the Raman effect will
only affect the evolution of the DSW as viscosity in the case of water waves dynamics
described by the KdV--B equation. Then, to consider the dynamics
of shock waves, it makes sense to confine ourselves to such initial conditions,
which evolution leads only to the DSW formation
without the formation of a rarefaction wave. This will greatly simplify the study.
In the Riemann invariants notation, this
means that the initial state should be chosen according to the
conditions \textit{(a)} $r_+^L = r_+^R$, $r_-^L > r_-^R$,
\textit{(b)} $r_+^L > r_+^R$, $r_-^L = r_-^R$,
where the superscript denotes the corresponding side of discontinuity.

Having defined the initial conditions in such a way,
let us turn to the description of the three stages of
the DSW formation with the Raman effect.

\subsection{First stage: $x\ll\gamma^{-1}$}

Since wave breaking occurs immediately, the dispersion
must be taken into account at this stage, when $x\ll\gamma^{-1}$.
However, the wave modulation is relatively large and the
Raman effect can be neglected. Then the dynamics is described by
the standard NLS equation (\ref{NLS_Dimentionless}) which periodic solution
is well known and can be written in the form
(see, for instance, \cite{kamch-2000})
\begin{equation} \label{Periodic_of_NLS}
\begin{split}
    I&=\frac{1}{4}(\lambda_4-\lambda_3-\lambda_2+\lambda_1)^2+\\
    &+(\lambda_4-\lambda_3)(\lambda_2-\lambda_1)\times\\
    &\times\mathrm{sn}^2(\sqrt{(\lambda_4-\lambda_2)(\lambda_3-\lambda_1)}\theta,m), \\
    u&=V-\frac{j}{I},
    \end{split}
    \end{equation}
where
\begin{equation} \label{theta_m_j_Equation}
\begin{split}
    & \theta=t-\frac{x}{V},\qquad \frac{1}{V}=\frac{1}{2}\sum_{i=1}^{4}\lambda_i,\\
    & m=\frac{(\lambda_2-\lambda_1)(\lambda_4-\lambda_3)}{(\lambda_4-\lambda_2)(\lambda_3-\lambda_1)},\quad 0\leq m\leq1; \\
    & j=\frac{1}{8}(-\lambda_1-\lambda_2+\lambda_3+\lambda_4)\times\\
    &\times(-\lambda_1+\lambda_2-\lambda_3+\lambda_4)
    (\lambda_1-\lambda_2-\lambda_3+\lambda_4);
\end{split}
\end{equation}
and the real parameters $\lambda_i$ are zeros of the polynomial
\begin{equation} \label{P_Equation}
    P(\lambda)=\prod_{i=1}^{4}(\lambda-\lambda_i)=\lambda^4-s_1\lambda^3+s_2\lambda^2-s_3\lambda+s_4,
\end{equation}
where
\begin{equation} \label{s_Equation}
\begin{split}
    & s_1=\sum_i\lambda_i,\quad s_2=\sum_{i<j}\lambda_i\lambda_j,
    \quad s_3=\sum_{i<j<k}\lambda_i\lambda_j\lambda_k, \\
    & s_4=\lambda_1\lambda_2\lambda_3\lambda_4,
\end{split}
\end{equation}
and $\lambda_i$ are ordered according to the inequalities
\begin{equation*}
    \lambda_1\leq\lambda_2\leq\lambda_3\leq\lambda_4.
\end{equation*}
As we see, the inverse phase velocity $V^{-1}$,
the amplitude $a=(\lambda_4-\lambda_3)(\lambda_2-\lambda_1)$,
the background density  $I_0=(\lambda_4-\lambda_3-\lambda_2+\lambda_1)^2/4$
through which the wave propagates,
and the wavelength
\begin{equation} \label{WaveLenghtT}
    T=\frac{2K(m)}{\sqrt{(\lambda_1-\lambda_3)(\lambda_2-\lambda_4)}}.
\end{equation}
are expressed in terms of these parameters
($K(m)$ is the complete elliptic integral of the first kind).
In a DSW they become
slow functions of $x$ and $t$.
The periodic solution written in the form (\ref{Periodic_of_NLS}) has the
advantage that the parameters $\lambda_i$ are Riemann invariants
of the Whitham modulation equations,
and their evolution is defined by Whitham
equations in a diagonal Riemann form
\begin{equation} \label{Withem_Eq}
    \frac{\partial\lambda_i}{\partial x}+\frac{1}{v_i(\lambda_1,\lambda_2,\lambda_3,\lambda_4)}
    \frac{\partial\lambda_i}{\partial t}=0,\quad i=1,2,3,4.
\end{equation}
Here, $v_i$ are the characteristic Whitham velocities:
\begin{equation} \label{Whitham-veloc}
\begin{split}
    & \frac{1}{v_1}= \frac{1}{2}\sum_{i=1}^{4}\lambda_i-\frac{(\lambda_4-\lambda_1)
    (\lambda_2-\lambda_1)K}{(\lambda_4-\lambda_1)K-(\lambda_4-\lambda_2)E}, \\
    & \frac{1}{v_2}= \frac{1}{2}\sum_{i=1}^{4}\lambda_i+\frac{(\lambda_3-\lambda_2)
    (\lambda_2-\lambda_1)K}{(\lambda_3-\lambda_2)K-(\lambda_3-\lambda_1)E}, \\
    & \frac{1}{v_3}= \frac{1}{2}\sum_{i=1}^{4}\lambda_i-\frac{(\lambda_4-\lambda_3)
    (\lambda_3-\lambda_2)K}{(\lambda_3-\lambda_2)K-(\lambda_4-\lambda_2)E}, \\
    & \frac{1}{v_4}= \frac{1}{2}\sum_{i=1}^{4}\lambda_i+\frac{(\lambda_4-\lambda_3)
    (\lambda_4-\lambda_1)K}{(\lambda_4-\lambda_1)K-(\lambda_3-\lambda_1)E},
\end{split}
\end{equation}
where $E=E(m)$ is the complete elliptic integral of
the second kind.

In the limit $m\rightarrow1$ ($\lambda_3\rightarrow\lambda_2$)
a traveling wave transforms into a soliton solution
on constant background:
\begin{equation}
\begin{split}
    &I=\frac{1}{4}(\lambda_4-\lambda_1)^2-
    \frac{(\lambda_4-\lambda_2)(\lambda_2-\lambda_1)}
    {\ch^2(\sqrt{(\lambda_4-\lambda_2)(\lambda_2-\lambda_1)}\theta)},\\
    &\theta=x-\frac{1}{2}(\lambda_1+2\lambda_2+\lambda_4)t.
\end{split}
\end{equation}
In the other (small-amplitude) limit
$m\rightarrow0$ ($\lambda_3\rightarrow\lambda_4$ or $\lambda_2\rightarrow\lambda_1$)
the wave amplitude approaches zero, while
the density takes its background value. Significantly,
the pair of Whitham velocities in these limits transforms
to the Riemann velocities of the dispersionless limit.
This means that the edges of the DSW
match the smooth solutions of the dispersionless hydrodynamic
approximation.

Since the initial
condition contains no parameters with the dimensions
of time, the modulation parameters depend only
on the self-similar variable $\tau=t/x$. Therefore,
Whitham equations are reduced to
\begin{equation} \label{Self-Sim}
    \frac{d\lambda_i}{d\tau}(v_i-\tau)=0,\quad i=1,2,3,4.
\end{equation}
Hence, it follows that only one Riemann invariant is
varying, while the three remaining ones must have
constant values.
Thus, knowing the input conditions, we can get the form of DSW for $x\ll\gamma^{-1}$.
Characteristic situations have been studied in
\cite{gk-87,eggk-95} and we will not dwell here on the details.

\subsection{Second stage: $x\sim\gamma^{-1}$}

At this stage, when $x\sim\gamma^{-1}$, effects associated with the
Raman term begin to play a significant role and to compete
with the modulation effect.
The dynamics of the wave is governed by the complete
NLS equation with the Raman term (\ref{NLS_Raman}).
The local waveform is still described by a periodic solution
(\ref{Periodic_of_NLS}), but the Whitham equations now become
non-uniform due to the appearance of the perturbation term.
In their derivation we use the method developed in \cite{Kamch-2004}.
It can be formulated as follows.
Let the evolution equations of some field variables $\varphi_m(x,t)$ have the
form
\begin{equation} \label{General_Pereturb}
\begin{split}
    \varepsilon\frac{\prt \varphi_m}{\prt t} & = K_m(\varphi_n,\varepsilon
    \frac{\prt \varphi_m}{\prt x},\varepsilon^2\frac{\prt^2 \varphi_m}{\prt x^2},...) \\
    & + R_m(\varphi_n,\varepsilon\frac{\prt \varphi_m}{\prt x},
    \varepsilon^2\frac{\prt^2 \varphi_m}{\prt x^2},...), \\
    & \qquad m,n=1,...,N,
\end{split}
\end{equation}
where a small parameter $\varepsilon\ll1$ measures the dispersion effects,
$K_m$ are the functions which correspond to the ``leading''
integrable part of the equations without taking into account
perturbations, and the functions $R_m$ are the perturbing terms
of the system. It is supposed that a nonperturbed
system can be represented as a compatibility condition of two linear
equations,
\begin{equation} \label{LinEqAB}
\begin{split}
    \varepsilon^2 \chi_{xx} & = \mathcal{A}\chi, \\
    \chi_t & = -\frac12 \mathcal{B}_x\chi+\mathcal{B}\chi_x,
\end{split}
\end{equation}
where $\mathcal{A}$ and $\mathcal{B}$ depend on the $\varphi_n$, their space derivatives,
and on the spectral parameter $\lambda$. This condition is satisfied
for the NLS equation and allows us to use
the powerful finite-gap integration method
for its investigation (see, for instance, \cite{kamch-2000}).
The second-order linear equations (\ref{LinEqAB}) have two basis solutions
$\chi_\pm$ and their product $g = \chi_+ \chi_-$ satisfies a third-order differential
equation which can be integrated once to give
\begin{equation} \label{}
\begin{split}
    \frac{\varepsilon^2}{2}gg_{xx}-\frac{\varepsilon^2}{4}g_x^2
    -\mathcal{A}g^2=\sigma P(\lambda),
\end{split}
\end{equation}
where $\sigma$ is determined by the sign of the highest order
term $\mathcal{A}$ as a function of $\lambda$
($\mathcal{A}|_{\lambda\rightarrow\infty}\rightarrow-\sigma\lambda^r$),
$P(\lambda)$ is polynomial in $\lambda$, and $\lambda_i$ are its zeros.

In a modulated wave the parameters $\lambda_i$ become slow functions of $x$ и $t$,
whose evolution is described by the Whitham equations which in the case of
(\ref{General_Pereturb}) can be
written in the form
\begin{equation} \label{}
\begin{split}
    \frac{\partial \lambda_i}{\partial t}-\frac{\left<\mathcal{B}/g\right>}
    {\left<1/g\right>} & \frac{\partial \lambda_i}{\partial x} = \\
    = \lim_{\varepsilon\rightarrow0} & \left\{ \frac{\sigma}
    {\left<1/g\right>\prod_{j \neq k}(\lambda_k-\lambda_j)} \times \right. \\
    & \left. \times \sum_{m=1}^{N}\sum_{l=0}^{A_m}\left< \frac{\partial \mathcal{A}}
    {\partial \varphi_m^{(l)}}\frac{\partial^l R_m}{\partial x^l} g \right> \right\}, \\
    & \qquad i=1,...,M,
\end{split}
\end{equation}
where $A_m$ denotes the highest order derivative of functions $\varphi_m$, entering
in $\mathcal{A}$.
The angle brackets denote the averaging over one
wavelength $L$. The spectral parameter $\lambda$ should be put equal to $\lambda_i$ after
averaging.

We shall apply here this scheme to the NLS equation with Raman term
\begin{equation} \label{eq3-ak}
    i\varepsilon q_x + \frac{\varepsilon^2}{2} q_{tt} - |q|^2q = - \varepsilon\gamma q(|q|^2)_t.
\end{equation}
The positive small parameter $\varepsilon\ll1$ introduced here will
not change the final equations, since in the limit
$\varepsilon\rightarrow0$ it disappears, leaving the leading
term of powers of $\varepsilon$.
In the equation (\ref{eq3-ak}) we have two field variables $q$ and $q^*$, and,
correspondingly, two terms of perturbation
\begin{equation} \label{}
    R_q=i\gamma q(|q|^2)_t, \qquad R_{q^*}=-i\gamma q^*(|q|^2)_t.
\end{equation}
For the nonperturbed NLS equation the functions $\mathcal{A}$ and $\mathcal{B}$ are
specified as
\begin{equation} \label{}
\begin{split}
    \mathcal{A} & = -\lambda^2 + i\varepsilon\lambda\frac{q_t}{q} + |q|^2 -
    \frac{\varepsilon^2}{2}\frac{q_{tt}}{q} + \frac{3\varepsilon^2}{4}\frac{q_t^2}{q^2}, \\
    \mathcal{B} & = -\lambda + \frac{i\varepsilon}{2}\frac{q_t}{q}.
\end{split}
\end{equation}
The averaging can be performed with the use of equations known
from the theory of periodic solutions of the NLS equation
\begin{equation} \label{}
\begin{split}
    g=\lambda-\mu, \qquad \varepsilon\frac{d\mu}{dt}=2\sqrt{-P(\mu)}, \\
    \frac{1}{V}=\frac{s_1}{2}, \qquad T = \varepsilon\oint\frac{d\mu}{2\sqrt{-P(\mu)}},
\end{split}
\end{equation}
where $P(\mu)$ and $s_1$ are defined by the equations (\ref{P_Equation}) and (\ref{s_Equation}).
To calculate the right-hand side parts of the Whitham equations,
we also need an expression for the parameter $\mu$ (see \cite{kamch-2000})
\begin{equation} \label{}
\begin{split}
    \mu(I) & = \frac{s_1}{4}+\frac{-j+i\sqrt{\mathcal{R}(I)}}{2I}, \qquad
    \varepsilon\frac{dI}{dt} = 2\sqrt{\mathcal{R}(I)}, \\
    \mathcal{R}(I) & = (I-I_1)(I-I_2)(I-I_3),
\end{split}
\end{equation}
where $I_i$ are the zeros of function $\mathcal{R}(I)$ related to Riemann invariants
$\lambda_i$ by the formulas
\begin{equation} \label{}
\begin{split}
    I_1=\frac14 (\lambda_1-\lambda_2-\lambda_3+\lambda_4)^2, \\
    I_2=\frac14 (\lambda_1-\lambda_2+\lambda_3-\lambda_4)^2, \\
    I_3=\frac14 (\lambda_1+\lambda_2-\lambda_3-\lambda_4)^2,
\end{split}
\end{equation}
and $j$ is given by the equation (\ref{theta_m_j_Equation}).
Then we obtain the Whitham equations for the Riemann
invariants $\lambda_i$ in the form,
\begin{equation} \label{Whitham_Perturbed_Full}
\begin{split}
    \frac{\partial \lambda_i}{\partial x} & +\frac{1}{v_i}\frac{\partial \lambda_i}{\partial t} = \\
        & = \frac{\gamma j}{\partial T/\partial \lambda_i}\frac{\lambda_i}{\prod_{j \neq i}(\lambda_i-\lambda_j)}
        \int_{I_1}^{I_2}\frac{\sqrt{\mathcal{R}(I)}}{I^2} dI, \\
        & \qquad\qquad i=1,2,3,4,
\end{split}
\end{equation}
where the wavelength $T$ is expressed by the equation (\ref{WaveLenghtT}).
The integral on the right-hand side can be expressed in terms of
elliptic integrals. However, this expression is very
complex, and it is easier to deal with its original non-integrated form.

The solution of the perturbed Whitham equations
(\ref{Whitham_Perturbed_Full}) determines the evolution
of the parameters $\lambda_i$ due to the non-uniform wave modulation
and the weak effect of the viscosity type. By the analogy with
studied earlier small-amplitude
limit, it is natural to expect that DSW
propagating in a positive direction will asymptotically tend
to a stationary wave. Some of its characteristics can be
found analytically. The amplitude of the DSWs propagating to the left will continuously increase.
This solution can be found numerically from the perturbed Whitham equations.

\subsection{Third stage: $x\gg\gamma^{-1}$}

At distances $x\gg\gamma^{-1}$ the DSW
propagating in the positive direction of the $t$-axis becomes stationary,
since for such $x$ the Raman effect will balance the wave modulation effect.
This means that DSW moves as a whole with constant velocity
$V$ and its profile does not change.
Then the DSWs will be determined by the perturbed Whitham equations
(\ref{Whitham_Perturbed_Full}) with the parameters $\lambda_i$ depending on $\theta=t-x/V$.
If we assume that the function $s_1$ is an integral of the Whitham equations,
that is $s_1=\mathrm{const}$ under the condition $1/V=s_1/2$,
then the Whitham equations take simpler form
\begin{equation} \label{Whitham_Perturbed_Stat}
    \frac{\partial \lambda_i}{\partial \theta}=
        \frac{\lambda_iQ}{\prod_{j \neq i}(\lambda_i-\lambda_j)}, \qquad i=1,2,3,4,
\end{equation}
where
\begin{equation} \label{}
\begin{split}
    Q = \frac{2\gamma j}{T} \int_{I_1}^{I_2}\frac{\sqrt{\mathcal{R}(I)}}{I^2} dI.
\end{split}
\end{equation}

\begin{figure}[t] \centering
\includegraphics[width=8cm]{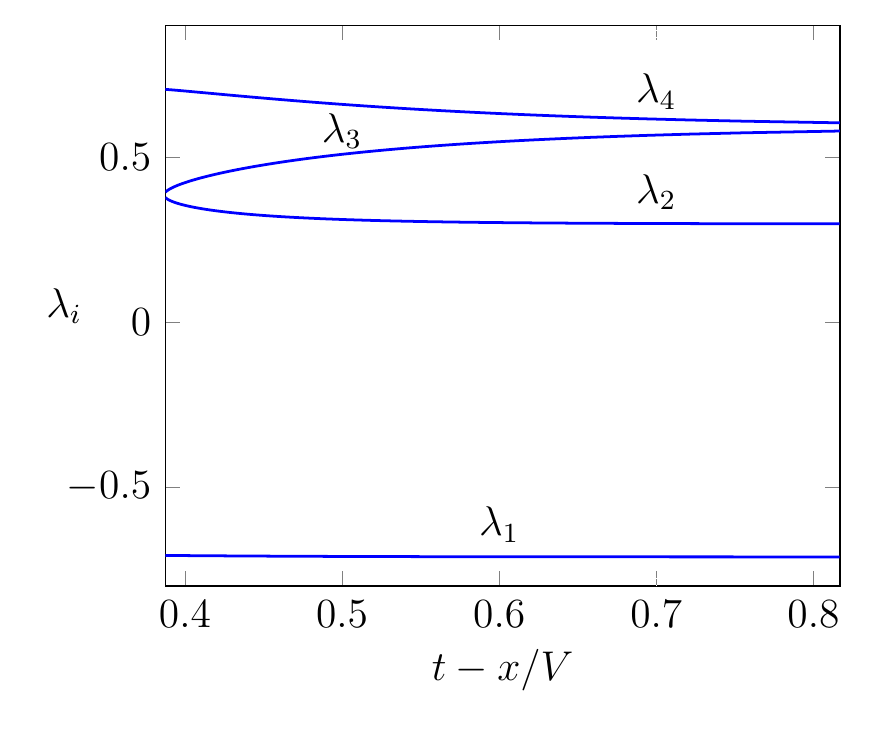}
\caption{Riemann invariants $\lambda_i$ as functions of $\theta=t-x/V$.}
\label{Fig3}
\end{figure}

We will show that the structure of these equations
actually provides three integrals $s_1$, $s_2$ and $s_4$.
This statement can be proved using Jacobi identities
which follow from the obvious identity
\begin{equation} \label{Jakobi}
    \sum_{i=1}^M\frac{\prod_{j \neq i}(\lambda-\lambda_j)}{\prod_{j \neq i}(\lambda_i-\lambda_j)} = 1,
\end{equation}
where on the left-hand side we have a polynomial of degree
$M-1$ which is equal to one at $N$ points
$\lambda=\lambda_i$, $I=1,...,M$ and therefore equals to one
identically. In our case $M=4$, so from this identity, we get
\begin{equation} \label{}
\begin{split}
    & \sum_{i=1}^4 \frac{\lambda_i^n}{\prod_{j \neq i}(\lambda_i-\lambda_j)} =
        \begin{cases}
            0, & \quad n = 0,1,2 \\
            1, & \quad n = 3
        \end{cases} , \\
    & \sum_{i=1}^4 \frac{\lambda_i^n{\sum_j}'\lambda_j}{\prod_{j \neq i}(\lambda_i-\lambda_j)} =
        \begin{cases}
            0, & \quad n = 0,1,3 \\
            -1, & \quad n = 2
        \end{cases} , \\
    & \sum_{i=1}^4 \frac{\lambda_i^n{\sum_{j,k}}'\lambda_j\lambda_k}{\prod_{j \neq i}(\lambda_i-\lambda_j)} =
        \begin{cases}
            0, & \quad n = 0,2,3 \\
            1, & \quad n = 1
        \end{cases} , \\
    & \sum_{i=1}^4 \frac{\lambda_i^ns_4}{\prod_{j \neq i}(\lambda_i-\lambda_j)} =
        \begin{cases}
            0, & \quad n = 0,1,2 \\
            -1, & \quad n = -1
        \end{cases} , \\
\end{split}
\end{equation}
where the sum with a prime means that all
terms with a factor $\lambda_i$ are omitted.
It follows that
\begin{equation} \label{}
    \frac{ds_1}{d\theta}  = 0, \quad
    \frac{ds_2}{d\theta} = 0, \quad
    \frac{ds_3}{d\theta}  = Q, \quad
    \frac{ds_4}{d\theta} = 0,
\end{equation}
that is, the equations (\ref{Whitham_Perturbed_Stat})
have three integrals of motion $s_1=\mathrm{const}$,
$s_2=\mathrm{const}$ and $s_4=\mathrm{const}$.
Thus, the system of equations (\ref{Whitham_Perturbed_Stat})
is reduced to a single ordinary differential equation
\begin{equation} \label{}
\begin{split}
    \frac{ds_3}{d\theta} = \frac{2\gamma j}{T} \int_{I_1(s_3)}^{I_2(s_3)}
    \frac{\sqrt{\mathcal{R}(I)}}{I^2} dI
\end{split}
\end{equation}
with the initial condition $s_3 (\theta_0)=s_3^0$,
which is more convenient to take at the soliton edge of the DSW.

\begin{figure}[t] \centering
\includegraphics[width=8cm]{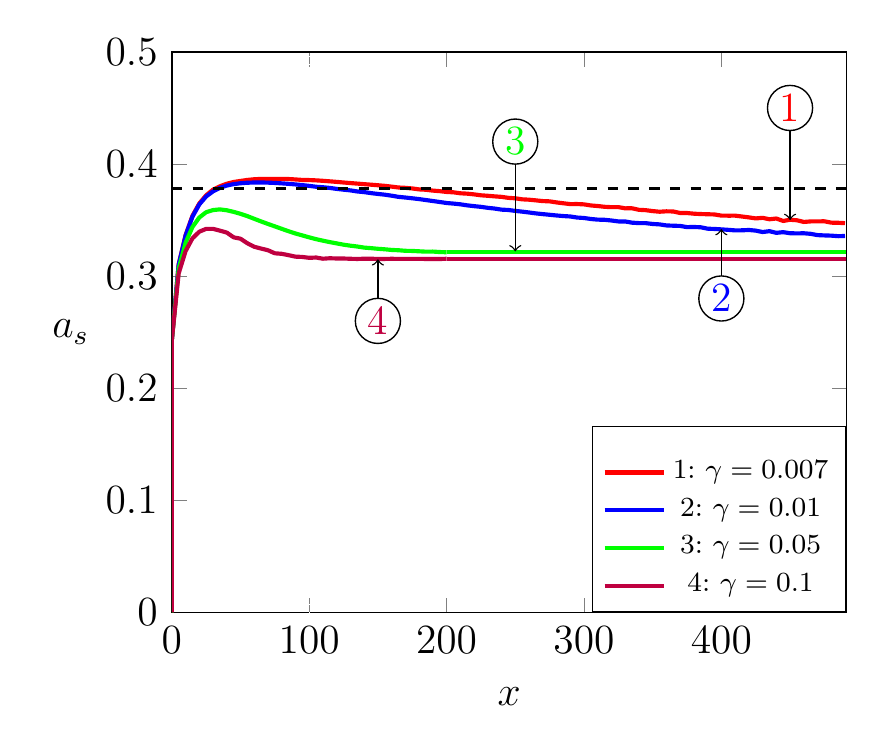}
\caption{The dependence of the amplitude of the leading soliton
on the propagation length $x$ at different values of the constant $\gamma$
characterizing the Raman effect. The analytical result
obtained with the help of the Whitham theory
is shown by a black dashed line, and numerical results obtained
by the solution of equation (\ref{NLS_Raman}) are shown by solid curves.}
\label{Fig4}
\end{figure}

\begin{figure}[t] \centering
\includegraphics[width=8cm]{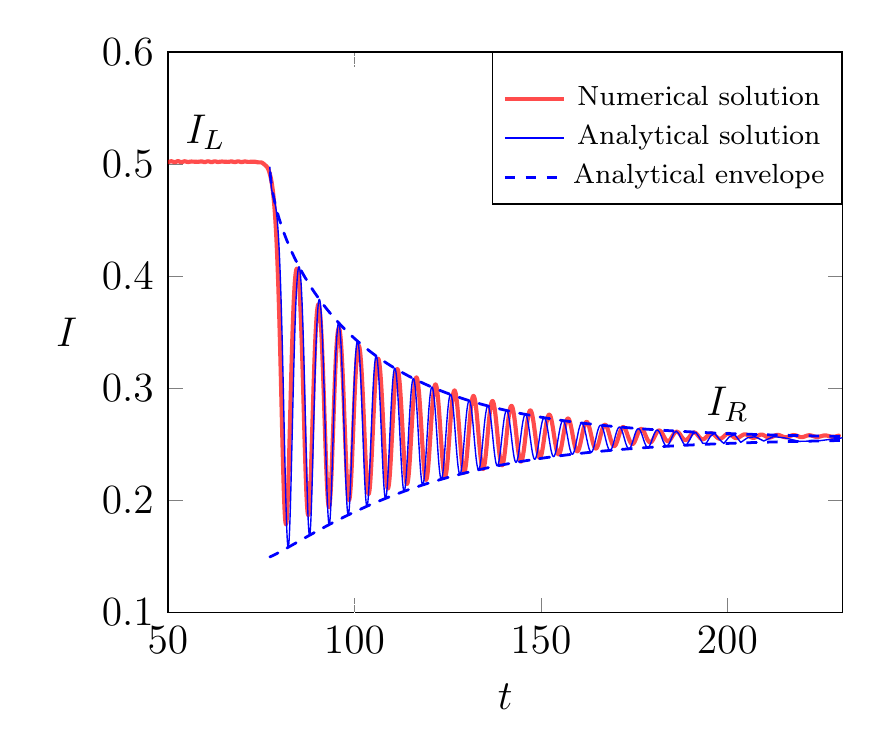}
\caption{Stationary DSW.
Comparison of the Whitham theory (blue color)
with numerical solution of the equation (\ref{NLS_Raman}) (red color)
for $\gamma=0.05$ with boundary conditions $I_L=0.5$, $I_r=0.257$,
$u_L=0$ and $u_R=-0.4$ at $x=200$ is shown.
The dashed lines show the results of the Whitham theory for the envelopes.}
\label{Fig5}
\end{figure}

From the matching condition at the edges of the DSW we find that at the soliton edge
\begin{equation} \label{}
  \lambda_1^L = r_-^L, \quad \lambda_4^L = r_-^L \quad \text{if} \quad \lambda_3^L=\lambda_2^L,
\end{equation}
and  at the small-amplitude edge
\begin{equation} \label{}
  \lambda_1^R = r_-^R, \quad \lambda_2^R=r_+^R \quad \text{if} \quad \lambda_3^R=\lambda_4^R.
\end{equation}
A diagram of Riemann invariants is shown in figure \ref{Fig3}.
Since in this case we consider a wave propagating in a positive direction,
we have $r_+^L > r_+^R$, $r_-^L = r_-^R$.
This means that $\lambda_4^L > \lambda_2^R$ and
$\lambda_1^L = \lambda_1^R$. Then from the constancy of
functions $s_1$ and $s_2$ we get
\begin{equation} \label{}
\begin{split}
  \lambda_2^L & = \frac{1}{2}\left(\lambda_2^R+\sqrt{\lambda_2^R\lambda_4^L}\right) \\
  & = \frac{1}{4}\left(u^R+2\sqrt{I^R}+\sqrt{(u^L+2\sqrt{I^L})(u^R+2\sqrt{I^R})}\right)
\end{split}
\end{equation}
and
\begin{equation} \label{}
\begin{split}
  \lambda_4^R & = \frac{1}{2}\left(\lambda_4^L+\sqrt{\lambda_2^R\lambda_4^L}\right) \\
  & = \frac{1}{4}\left(u^L+2\sqrt{I^L}+\sqrt{(u^L+2\sqrt{I^L})(u^R+2\sqrt{I^R})}\right).
\end{split}
\end{equation}
Thus, we know all Riemann invariants on both edges of DSW,
that is, we know integrals $s_1$ and $s_4$ along the whole shock wave.
From here we can get the wave velocity and amplitude of the leading soliton
\begin{equation} \label{}
\begin{split}
  \frac{1}{V} & = \frac{s_1}{2}=\frac{1}{2}\left( \lambda_1^L+\lambda_4^L
  +\lambda_2^R+\sqrt{\lambda_2^R\lambda_4^L} \right) \\
  & = \frac{1}{4}\left( 2u^L+u^R+2\sqrt{I^R} \right.\\
  & \left. \qquad + \sqrt{(u^L+2\sqrt{I^L})(u^R+2\sqrt{I^R})} \right),
\end{split}
\end{equation}
\begin{equation} \label{}
\begin{split}
  a_s & = \frac{1}{4}\left( 2\lambda_4^L-\lambda_2^R-\sqrt{\lambda_2^R\lambda_4^L} \right)
  \left( \lambda_2^R-2\lambda_1^L+\sqrt{\lambda_2^R\lambda_4^L} \right) \\
  & =  I^L - \frac{1}{16}\left[ 2\sqrt{I^L}-2u^L+u^R  \right.\\
  & \left. \qquad +\sqrt{(u^L+2\sqrt{I^L})(u^R+2\sqrt{I^R})} \right]^2.
\end{split}
\end{equation}
It should be noted that the velocity of the DSW
and soliton amplitude depend only on the
initial parameters and do not depend on the constant
$\gamma$, which reflects the influence of the Raman effect.
The numerical results reflecting the dependence of the soliton
amplitude on the coordinate $x$ are shown in figure \ref{Fig4}.
As we can see, there is some deviation of the analytical theory from
the numerical calculation, apparently caused by the fact that the
Whitham theory does not take into account non-adiabatic effects.
It may play a significant role in DSWs theory described by non-integrable
equations. Such differences have already been noted in the works
\cite{egs-06, egkkk-07} for other physical systems.
Despite this difference, DSW is well described
by the Whitham theory, as is illustrated in figure \ref{Fig5}.
As one can see, the correspondence between the results of Whitham
theory and numerical calculations improves noticeably with
decrease of the wave amplitude.

\begin{figure}[t] \centering
\includegraphics[width=8cm]{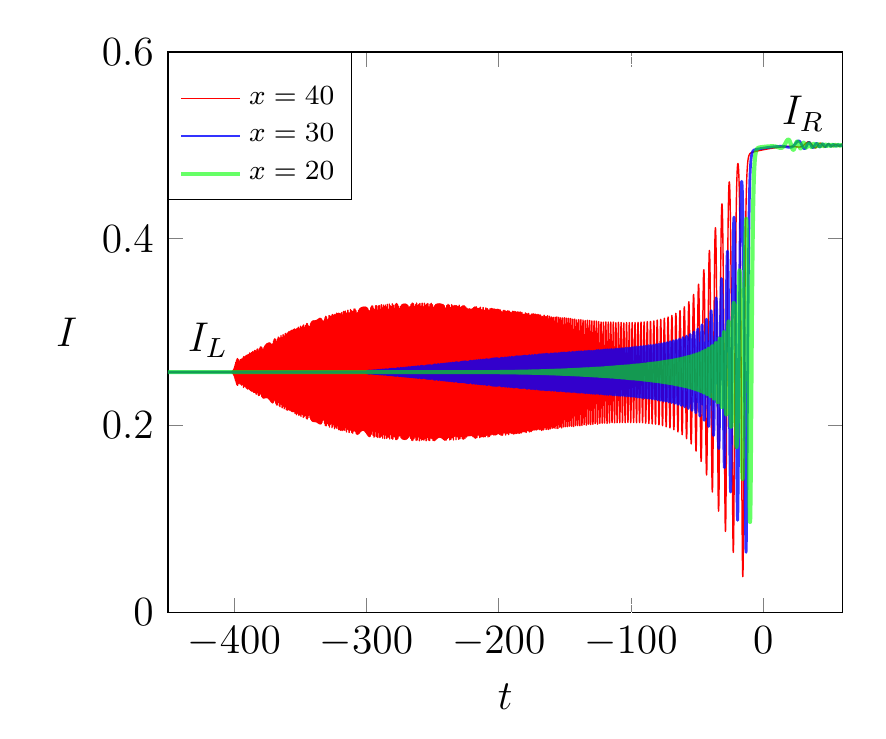}
\caption{Dispersive shock wave propagating in the negative
direction of the $t$-axis for $\gamma=0.1$ with boundary
conditions $I_l=0.257$, $I_r=0.5$, $u_L=0.4$ and $u_R=0$.}
\label{Fig6}
\end{figure}

For a wave which propagates in the opposite direction,
the reverse situation will occur. Amplitude of
oscillations in the wave and its length will continuously increase.
The construction of an analytical theory of such a non-stationary
nonlinear structure is very difficult and one has to turn to
numerical calculation. Figure \ref{Fig6} shows an example of
such a wave obtained by numerical solution of the equation (\ref{NLS_Raman}).
It can be seen that the wave profile is significantly different
for waves at different distances $x$. In this case, at
sufficiently large waveguide lengths $x$, one can see a
significant increase in the amplitude of the wave packet
separated from the DSW. Description of such a wave packet obeys
with good accuracy the linear theory from section \ref{sec3}.

\section{Conclusion}

In this work, the propagation of sufficiently long pulses
in fibers described by the nonlinear Schr\"odinger equation
modified by a small term which characterizes the Raman effect was studied analytically.
The main stages of a shock wave formation with an
initial profile in the step-like form are considered.
Analytical solutions are constructed for the initial
unsteady and final stable states using the Whitham method.
Perturbed Whitham integrable equations for a nonlinear
Schr\"odinger equation with the Raman term were obtained
using the finite-gap integration method.

In principle, one may hope that the results found here
can be observed experimentally in systems similar to that
used in the recent experiment \cite{Xu-2017}.
However, one should keep in mind that in standard fibers
in addition to the Raman effect,
self-steepening effect also occurs.
However, the manifestations of these two effects are quite
different and therefore they can be identified separately.
As shown in the article \cite{IvKamch-2017},
the main consequence of self-steepening is the formation of
combined shock waves caused by the non-monotonic dependence
of the nonlinear term on the wave amplitude, while Raman
scattering leads to the formation of stationary shock waves at finite length.
At the same time, the Raman effect is usually much stronger
than the self-steepening effect. The theory developed in this article
shows that the Whitham method provides a general efficient approach
for description DSWs in fibers and other
optical systems.

\section*{Conflict of interest}

The authors declare that there is no conflict of interest.

\end{document}